\documentstyle[]{l-aa}

\def\vecr{{\bf r}}
\def\vecv{{\bf v}}

\newcommand{\fr}[2]{\frac{\displaystyle #1}{\displaystyle #2}}

\newcommand{\der}[3]{\fr{d^{#3} {#1}}{d {#2}^{#3}}}

\newcommand{\vphi}{\varphi}
\newcommand{\veps}{\varepsilon}

\newcommand{\Op}{\Omega_p}

\begin{document}
\title{Numerical simulation of large-scale magnetic-field evolution in
spiral galaxies}
%
 \author{ K.~Otmianowska-Mazur\inst{1,2} and
          M.~Chiba\inst{2,3}}
 \offprints{ K.~Otmianowska-Mazur (Krakow), e-mail: otmian@oa.uj.edu.pl}
 \institute{ Astronomical Observatory, Jagiellonian University,
             ul. Orla 171, Krakow, Poland
 \and        Max-Planck-Institut f\"ur Radioastronomie, Auf dem H\"ugel
             69, D-53121 Bonn, Germany
 \and        Astronomical Institute, Tohoku University,
             Sendai 980-77, Japan}
 \thesaurus{02.13.1; 02.13.2; 11.13.2; 11.11.1; 11.19.2}
 \date{Received date 4 August 1994; accepted date 11 February 1995}
%
\maketitle
\markboth{K.~Otmianowska-Mazur \& M.~Chiba: Galactic magnetic fields}{}


 \begin{abstract}
The evolution of large-scale magnetic fields in disk galaxies is
investigated numerically. The gasdynamical simulations in a disk
perturbed by spiral or bar potential are incorporated into the
kinematic calculations of induction equations to elucidate the effects
of non-axisymmetric disk structure on magnetic fields.
The effects of interstellar turbulence are given as the
turbulent diffusion of magnetic fields.
The usually adopted dynamo mechanism of
$\alpha$-effect is not considered in our
computations, because it is not obvious about the actual existence
of the effect in a galaxy.
Our principal concern
is to clear
how {\em observationally} and {\em theoretically well-established}
gas flow affects the magnetic-field structure and evolution,
without putting a lot of artificial parameters in the model.

We have found that the density-wave streaming motion of
gas has a significant
influence on the distribution of magnetic fields: the lines of
force are well aligned with spiral arms due to the compressional
and additional shearing flow of gas in these regions. In the inter-arm regions,
the field lines have the finite angles with respect to the imposed arms,
because the gaseous streamlines induced by spiral arms deviate from the
orientation of spiral arms themselves. These properties of magnetic-field
orientation across the arm are well in agreement with the results of
radio continuum observations. All simulation models have resulted in
the eventual decay of magnetic energy due to the strong turbulent diffusion.

We have also explored the azimuthally periodic function for the coefficient
of turbulent diffusion with the maximum in spiral-arm regions. This is
anticipated from the enhancement of turbulence by young OB stars and
supernovae. This effect of non-uniform dissipation gives rise to
the modulation of magnetic-field structure and its time evolution.
In particular, the dissipation rate of magnetic fields can be much smaller
than the usually assumed rate of $\sim 1/10^8$ yr$^{-1}$.


 \end{abstract}
%
\section{Introduction}

Measurements of linear polarization
in radio continuum observations have revealed characteristic
morphology of magnetic field configuration in galaxies
(Beck 1993): lines of force are well
aligned with characteristic features of a disk galaxy
such as spiral arms and bars,
and a few but not all edge-on galaxies prevail largely vertical fields
perpendicular to a disk. Attempts to derive actual directions of
magnetic vectors are rather delicated tasks with a lot of
ambiguities, and thus have not well
succeeded to uncover whether field directions change along the azimuth
or not in a galactic disk.
The recent high-resolution measurements
also allow to derive the fine structure
of the large-scale magnetic fields in a disk: pitch angles
of these fields vary with azimuthal angle in the systematic way
(e.g. Ehle \& Beck 1993),
and the well uniform fields are most recognizable
outside the optical spiral arms.
Apparently, these observations suggest that origin of
magnetic morphology is principally associated with a non-axisymmetric
galactic structure accompanying a characteristic large-scale gas flow.

It has been argued that present structure and typical strength
of galactic magnetic fields are organized by induction
processes called galactic $\alpha\omega$-dynamos (Krause et al. 1993).
Galactic dynamos are invoked by the self-amplification of magnetic fields
via helical small-scale turbulence ($\alpha$-effects) and differential
galactic rotation ($\omega$-effects). Generally initial (seed) magnetic fields
are composed of several magnetic modes, and the modes with finite
exponential growth are supposed to predominate at the present epoch
after some nonlinear saturation of growth.
The conventional models are characterized by a rather slow time scale of
field amplification, of the order of Gyr, and attempts have been made
to explain the present field structure, mainly based on an axisymmetric
disk model with some additional effects of vertical flow.
An implicit assumption in such an approach
is that a galactic structure is only
the second order effect on a magnetic-field structure, and dynamo
experts tend to invoke the properties of turbulence (such as
tensor $\alpha$) to explain non-axisymmetric magnetic fields
(e.g. R\"udiger et al. 1993).

Basically an induction equation says that the magnetic-field evolution
is strongly governed by the profiles of velocity fields, which are
typically having a time scale for change, of the order of 10$^8$ yrs,
in a galactic disk. Then it is followed that most of the present field
structures reflect the velocity fields of gas which have happened within
$\sim 10^8$ yrs, irrespective of slow $\alpha\omega$-dynamo processes.
Dynamos might give a solution for global field geometry (e.g. axisymmetric
or bisymmetric), which is however not well defined from observations.
It has also been claimed that dynamos cannot explain the strong field
at high-redshift intergalactic clouds (Wolfe et al. 1992).
Therefore, as most observational results suggest as mentioned above,
it is {\it essential} to clarify the role of a galactic structure in
understanding the {\it present} structure of magnetic fields
(see Kronberg 1994 and Chiba \& Lesch 1994
for more detailed discussion about the present status).

In the present paper, we focus on the evolution of magnetic fields
under the influence of non-axisymmetric disturbance, such as
spirals and bars, and perform numerical simulations to elucidate
how such disturbances affect the evolution and structure of
magnetic fields. With the same motivation, Panesar \& Nelson (1992)
have shown from their simulation that spiral arms strongly modify
the field morphology. Since they have performed only one case
due to the extensive nature of simulation, we attempt to consider
several cases for the parameters of disturbances, particularly concerning
the dependence on pitch angles of spiral arms, and spatial profiles
of turbulent diffusion for magnetic fields.

The plan of the paper is as follows:
Sec.2 is devoted to methods of simulations and model parameters.
Results of simulation are explained in Sec.3. Conclusions and discussion,
especially the comparison with observations, are given in Sec.4.

\section{Numerical Methods and Model parameters}

\subsection{Hydrodynamics in a galaxy}
In our simulations of magnetic fields in galaxies, it is crucial
to obtain the realistic large-scale flow of gas, which is
most plausible theoretically and also reminiscent of observed
velocity fields of the neutral gas component as well as the warm
ionized gas. We adopt the hydro-code of smoothed particle
hydrodynamics (SPH), first introduced by Lucy (1977) and
Gingold \& Monaghan (1977). The SPH code is based on a particle
method, without defining a computational grid. Thus it is fully
Lagrangian, and is naturally implemented in three-dimensional
calculations. Because of its flexible nature of the scheme,
it is widely used for many astrophysical problems
(see e.g. Monaghan 1992; Benz 1990; Hernquist \& Katz 1989).

\subsubsection{Fundamentals in SPH}
In SPH, the computation consists of a number of discrete fluid
elements, i.e. particles, so that some local averages must be
performed using an interpolating kernel $W(r;h)$. $h$ is the
smoothing length. Then any mean value of physical quantity,
say $A(\vecr)$ can be defined by
\begin{equation}
<A(\vecr)> = \int A(\vecr') W(\vecr - \vecr' ; h) d\vecr',
\label{inte}
\end{equation}
where the kernel $W$ is normalized, $\int W(\vecr;h) d\vecr = 1$.
When the values of $A(\vecr)$ are known only at finite and
discrete points, the integral interpolant is replaced by
the following equation,
\begin{equation}
<A(\vecr)> = \sum_{j=1}^N m_j \fr{A(\vecr_j)}{\rho_j}
              W(\vecr - \vecr_j ; h),
\label{sum}
\end{equation}
where $m_j$ is the particle mass; the density $\rho_j$ at each
particle's position is defined as
\begin{equation}
\rho = \sum_{j=1}^N m_j
              W(\vecr - \vecr_j ; h).
\label{rho}
\end{equation}
In our simulation, we adopt the kernel based on spline functions
(Monaghan \& Lattanzio 1985),

\begin{equation}
W(r;h) = \fr{1}{\pi h^3} \left\{
 \begin{array}{ll}
    1-(3/2)x^2 + (3/4)x^3  & {\rm for}\ 0 \leq x \leq 1 \\
    (1/4)(2-x)^3           & {\rm for}\ 1 \leq x \leq 2 \\
    0                      & {\rm for}\ x \geq 2,
 \end{array} \right.
\label{kern}
\end{equation}
where $x = r/h$. Here the smoothing length is variable in each particle,
$h_j$, according to the density at the particle's position (see below).

The hydrodynamical equations in our model are given by
\begin{eqnarray}
\der{\vecr_i}{t}{} &=& \vecv_i,
\label{hydeq1} \\
\der{\vecv_i}{t}{} &=& - \fr{1}{\rho_i} \nabla P_i + {\bf q}_i
                     - \nabla \Phi - 2\Op\times\vecv
                     - \Op\times\Op\times\vecr,
\label{hydeq2} \\
\der{h_i}{t}{} &=& - \fr{h_i}{3\rho_i} \der{\rho_i}{t}{},
\label{hydeq3}
\end{eqnarray}
where $P_i$ is the pressure, $\Phi$ the gravitational potential,
and ${\bf q}_i$ is the artificial viscosity to treat the
formation of shock waves. Since our aim is to obtain the characteristic
structure of velocity fields in a galaxy, where the effect of
self-gravity of gas is minor (except at the galactic center),
we neglect the contribution of gas itself in $\Phi$ but
consider the external galactic potential provided by galaxy mass
as a whole. $\Op$ denotes the pattern speed of non-axisymmetric
disturbance in a galactic structure (see next section). Therefore
we simulate the gas dynamics in a frame rotating with $\Op$,
where such disturbance is observed to be stationary.

The terms of pressure gradient and artificial viscosity are
written as:
\begin{eqnarray}
\fr{\nabla P_i}{\rho_i} + {\bf q}_i &=&
 - \sum_{j=1}^N m_j (\fr{P_j}{\rho_j^2}+\fr{P_i}{\rho_i^2}+\Pi_{ij})
\nonumber \\
  & & \times \fr{1}{2} \left[ \nabla_i W(r_{ij};h_i) +
      \nabla_i W(r_{ij};h_j) \right],
\label{pre}
\end{eqnarray}
where $\Pi_{ij}$ denotes the contribution of artificial viscosity
given by
\begin{equation}
\Pi_{ij} = \left\{
  \begin{array}{ll}
 \fr{-\alpha \bar{c}_{ij} \mu_{ij} + \beta \mu_{ij}^2}{\bar\rho_{ij}} f
& {\rm for} \ \vecv_{ij}\cdot\vecr_{ij} < 0 \\
 0 & {\rm for} \ \vecv_{ij}\cdot\vecr_{ij} \geq 0,
 \end{array} \right.
\label{vis1}
\end{equation}
and
\begin{equation}
\mu_{ij} = \fr{h_{ij} \vecv_{ij}\cdot\vecr_{ij} }{ \vecr_{ij}^2
+ \eta^2}.
\label{vis2}
\end{equation}
Here, $h_{ij}=(h_i + h_j)/2$, $\bar\rho_{ij} = (\rho_i+\rho_j)/2$,
and the average sound velocity, $\bar{c}_{ij} = (c_i+c_j)/2$.
In the expression for $\Pi_{ij}$, eq.(\ref{vis1}),
the first term corresponds to
a bulk viscosity while the second term is equivalent to the
Von Neumann-Richtmyer viscosity. For the parameters
($\alpha,\beta,\eta$), we adopt, $\alpha=1$, $\beta=2$,
and $\eta^2 = 0.01 h_{ij}^2$, which reproduce reasonably
some test calculations such as one-dimensional shock-tube
problem and two-dimensional wind-tunnel problem. The factor
$f$ is introduced to prevent the excessive shear viscosity
that appears in the standard form of viscosity given in
equs.(\ref{vis1}) and (\ref{vis2}). In our computation, we
use the form for $f$ described by Benz (1990) as,
\begin{equation}
f = \fr{|<\nabla\cdot\vecv>|}{|<\nabla\cdot\vecv>|+|<\nabla\times\vecv>|
+0.001 \bar{c}_{ij}/h_{ij} },
\label{swit}
\end{equation}
which vanishes if the quantity $<\nabla\times\vecv>$, the curl
of velocity is large, suppressing the shear viscosity.

The interstellar gas is modeled in terms of
the isothermal equations of state, $P=c^2 \rho$.

The basic equations eq.(\ref{hydeq1}), eq.(\ref{hydeq2}), and
eq.(\ref{hydeq3}) are integrated in time, using the leapfrog
method. The time step $\Delta t$ is controlled by the Courant
condition:
\begin{equation}
\Delta t = 0.3 min \fr{h_i}{h_i|\nabla\cdot\vecv|+c_i+1.2(\alpha c_i
+ \beta max_j |\mu_{ij}|)}.
\label{tstep}
\end{equation}

\subsubsection{Galaxy models}
The axisymmetric part of the gravitational potential in a disk plane
is represented as the Toomre disk (Toomre 1963) :
\begin{equation}
\phi_0(r) = - \fr{c^2}{a} (r^2+a^2)^{-1/2},
\label{phi0}
\end{equation}
which gives the maximum rotational velocity $v_{max}=(4/27)^{1/4} c/a$
at a radius $r_{max} = a \sqrt2$. For the vertical dependence of
the potential, we adopt a parabolic form, $\nu z^2 / 2$, where
$\nu$ is the vertical frequency for motions of stars. We use
the characteristic value in the solar neighborhood,
$\nu= 3.2 \times 10^{-15}$ s$^{-1}$ (Binney \& Tremaine 1987),
for all radii. Of course, this value
as well as the parabolic approximation does not hold all over
a disk, especially at high $z$ where this approximation is invalid.
The gas flow derived from this galactic potential is however
well representative of actual flow for the present purpose to
simulate the magnetic-field evolution.

Initially the gas disk has the approximately uniform density with
a finite thickness as determined by the vertical dependence
of the potential given above and
the sound speed of gas, $c$,
which is assumed to be constant, 8kms$^{-1}$. The disk is set up in
centrifugal equilibrium. The parameter $a$ in eq.(\ref{phi0}) is
2.6kpc in our simulation, and $c$ is determined so that $v_{max}=
200$ kms$^{-1}$kpc$^{-1}$.

Non-axisymmetric disturbances are introduced to a disk in
the first $1.6 \times 10^8$ yrs of time evolution, with the form,
\begin{equation}
\phi_1(r,\vphi) = \phi_0(r)\veps(r) \cos[2\ln(r/r_0)\cot \psi + 2\vphi],
\label{phi1}
\end{equation}
and are fixed hereafter (Matsuda \& Isaka 1980; Johns \& Nelson 1986).
$\psi$ is the pitch angle of spiral disturbance with a logarithmic
two-armed form, and $r_0$, arbitrary parameter,
determines the phase of spiral. $\veps(r)$ is represented as
\begin{equation}
\veps(r) = \veps_0 \fr{r^2}{a^2} ( 1 + \fr{r^2}{a^2} )^{-3/2},
\label{veps}
\end{equation}
which reaches the peak value $\veps_0 \sqrt{4/27}$ at a radius
$r_{max}$. In the simulation for a bar perturbation,
the term proportional to $\cot\psi$ in eq.(\ref{phi1}) is omitted.

Figure 1 shows the rotation curve and some frequency distribution
in our model. It follows that the strength of velocity shear
is strongest at $r \sim 3$kpc.

In order to obtain the characteristic velocity fields for solving the
magnetic-field evolution, the computations of gas dynamics are halted
after a few hundreds Myr when the gas flow becomes approximately steady,
and then the velocity fields are evaluated as will be explained in the
next section. We note that a small-scale time-dependent modulation
in gas dynamics is observed (cf. Johns \& Nelson 1986), but the obtained
velocity profiles by the above procedure are enough characteristic to
see their effects on magnetic fields.
As an example, the velocity-field vectors for the pitch angle 20$^{\circ}$
is presented in Fig.~2.
The profiles of gas density
for the spiral pitch angles, 10$^{\circ}$
and 20$^{\circ}$, and the bar perturbation
are shown in Fig.~3. The solid lines mark the minima of the gravitational
potential of spirals.

\begin{figure}[t]
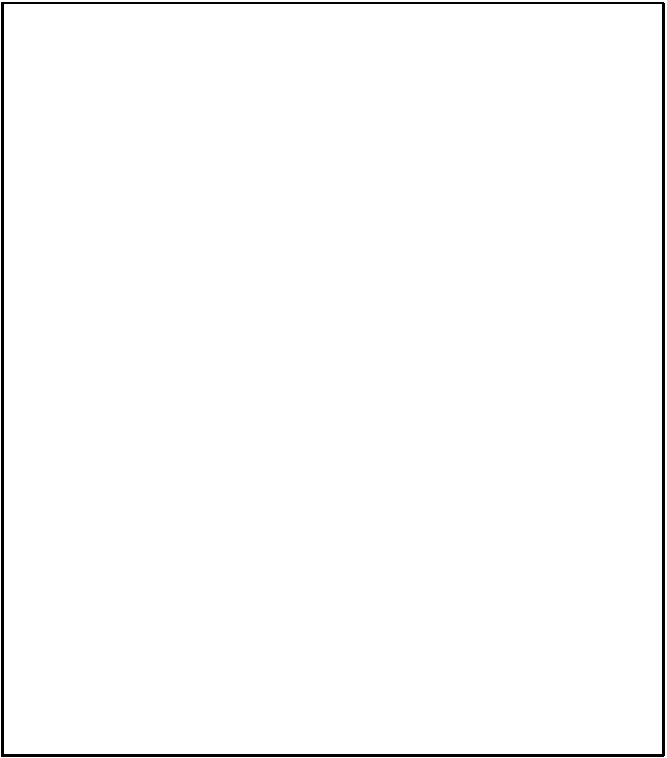

 \picplace{10.0 cm}
\caption[]{\label{fig1} The rotation curve (a) and frequency distribution (b)
  in the galaxy model}
\end{figure}
\begin{figure}[t]
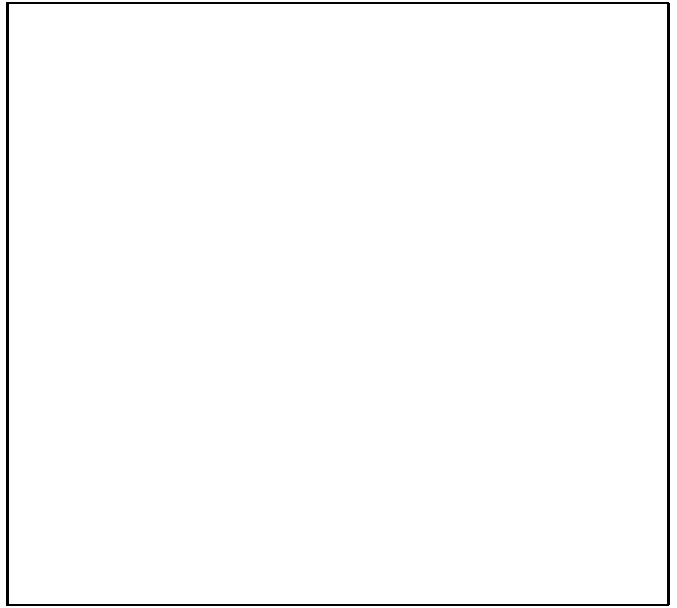

 \picplace{8.0 cm}
\caption[]{\label{fig2} The velocity field vectors in the case of
 the spiral pitch
 angle $\psi = 20^{\circ}$}
\end{figure}
\begin{figure}[hbtp]
 \picplace{22.0 cm}
\caption[]{\label{fig3} The profiles of gas density (grayplot)
in the cases of the spiral pitch angle $\psi = 10^{\circ}$
(a), $\psi = 20^{\circ}$ (b), and the bar (c). The solid lines
denote the minima of the gravitational potential of spirals}
\end{figure}

\subsection{Solving induction equations}

The kinematic evolution of the magnetic field is  analyzed  using  the  time
dependent solution of the field transport equation with  diffusion
(the induction equation),
\begin{equation}
\partial {\bf B}/\partial t~=~\nabla ~\times ~({\bf v~\times ~B})~-
\nabla ~\times \eta ~(\nabla ~\times {\bf B}),
\end{equation}
where $\eta $ is the diffusion coefficient, ${\bf B}$ and ${\bf v}$
are the magnetic and
velocity fields, respectively. The assumed approach allows us to
investigate qualitatively the effects of spatially variable $\eta$
on the evolution of magnetic fields in a disk. However,
the more precise solution of this problem is also possible
(Kichatinov \&
R\"udiger 1992).

The problem is solved numerically in three-dimensional $(3D)$
rectangular grid of points, where the $XY$  plane  is  the  galactic
plane and the $Z$-axis is the axis of galactic rotation pointing to the
North. The rectangular size in X and Y directions is
20~kpc, where the grid
interval is 400~pc. Along the Z axis the box has  3 or 4~kpc
depending on the assumptions concerning the diffusion coefficient (see below)
and with the spatial step of 250~pc interval.
The time interval is $\Delta t~=~0.22~\cdot~10^6$~yrs.
The fourth order explicit Runge-Kutta method  is  used  for
the discretization in time. The space derivatives are approximated
by the spline functions  of  the  third  order.  As  the  boundary
conditions, the continuity of the third order  derivatives  of  the
spline  functions  has  been  used.  Physically  it  means   that
impenetrable  walls  have  been  applied.  This  fact  causes  the
necessity of the velocity field decreasing near the edges.

The prevalent number of numerical  methods  applied  to  the
induction equation do not satisfy the divergence-free condition of
the magnetic field. There are several approaches  which  can  be
used to solve this difficult problem (Elstner et al. 1990;  Evans
\& Hawley 1988; Schmidt-Voigt 1989).  We  have adopted the
``divergence-cleaning'' method elaborated by Schmidt-Voigt (1989)
which   is
suitable in our case (see discussion in  Schmidt-Voigt 1989  and
Evans \& Hawley 1988). Following Schmidt-Voigt, in order  to
diminish  the  errors  of  the  non-zero  magnetic  divergency, we
introduce a  potential $\Phi_b$ which is  a solution of the Poisson
equation,
\begin{equation}
\nabla ^{2}~\Phi_b (x,y,z,t)~=~-(\nabla ~\cdot ~{\bf B}).
\end{equation}
After each time step, this  equation  is   solved   using   the
Gauss-Seidel iteration, and we replace the magnetic field ${\bf B}$ by
\begin{equation}
{\bf B}'= {\bf B~}+~\nabla ~\Phi_b.
\end{equation}
The iteration ends when the following condition is fulfilled:
\begin{equation}
\left|\matrix{(\nabla \cdot {\bf B})}\right|~\cdot ~\left|\matrix{\Delta x}
\right|/~\left|\matrix{{\bf B}}\right|~<~0.01.
\end{equation}
This is what is necessary to avoid the   unphysical    solutions
(Schmidt-Voigt 1987).

To incorporate the particle-based flow in SPH into the grid-based code
for magnetic-field evolution described above, the averaging of
velocity fields in grids is performed using the smoothing length
$h_i$ in SPH and Gaussian kernel $W_g$ (Habe, {\it et al.} 1991):
\begin{equation}
\vecv_{ijk} = \fr{1}{\rho_{ijk}} \sum_{n=1}^N m_n \vecv_n
               W_g(\Delta r_{ijk,n}; h_n),
\label{smooth}
\end{equation}
where $\Delta r_{ijk,n} = [(x_n-x_i)^2+(y_n-y_j)^2+(z_n-z_k)^2]^{1/2}$,
$\rho_{ijk}=\sum_{n} m_n W_g(\Delta r_{ijk,n}; h_n)$, and
$W_g = 1/\pi^{3/2}/h_n^3 \exp[-(r/h_n)^2]$. Here the indexes $i,j,k$
are for grids, and $n$ for particles. The reason to adopt the
Gaussian kernel
in this procedure is that due to the finite
number of particles ($N=2000$), there would be the lack of
velocities at some grids in the spline case, because $W$ defined
in eq.(\ref{kern}) is zero for $r/h > 2$. However, the different
choice of kernel for velocity fields at grids does not introduce
the fundamental change in the characteristic properties of gas
flow obtained in SPH.

The computations have been performed using
Cray~Y-MP in HLRZ J\"ulich and
SPARC~10   in   the
Max-Planck-Institut f\"ur Radioastronomie in Bonn.

\begin{table*}
\caption[]{The model parameters}
\begin{flushleft}
\begin{tabular}{lllllllll} \hline\noalign{\smallskip}
Model & $\veps_0$ & Pitch angle $\psi$    & $\Op$      & Corotation
  & $\eta$ & $B^{t=0}$(Inclination $i^{\circ}$) & $Z_{max}$ & Comments \\
      &           & ($^\circ$)            & (km/s/kpc) & (kpc)
  &        &                                    &       &      \\
\noalign{\smallskip}\hline\noalign{\smallskip}
A1    &  0.1      &  10                   &  18        & 9.1
    & $\eta(z)$ & ring(0)      & 3  &  $R_m=55$ \\
A2    &  0.1      & 10                    & 18         & 9.1
    & $\eta(r,\vphi,z)$  & ring(0)    & 4  &  $R_m=145$  \\
A3    &  0.1      & 20               & 18         & 9.1
    & $\eta(z)$ & ring(0)    & 3 &  $R_m=55$  \\
A4    &  0.1      & 90(bar)               & 30        & 6.2
    & $\eta(z)$ & ring(0)       & 3 &  $R_m=55$ \\
\noalign{\smallskip}\hline\noalign{\smallskip}
B1    &  0.0      & ---                   & ---        & ---
    & $\eta(z)$ & uniform(0)   & 3 & circular rotation, $R_m=55$      \\
B2    &  0.1      & 10                    & 18         & 9.1
    & $\eta(z)$ & uniform(0)   & 3 &  $R_m=55$  \\
B3    &  0.1      & 20                    & 18         & 9.1
    & $\eta(z)$ & uniform(0)   & 3 &  $R_m=55$  \\
B4    &  0.1      & 90(bar)               & 30         & 6.2
    & $\eta(z)$ & uniform(0)   & 3  & $R_m=55$\\
B5    &  0.1      & 10                    & 18         & 9.1
    & $\eta(r,\vphi)$ & uniform(0)   & 3 & $R_m=178$  \\
B6    &  0.1      & 10                    & 18         & 9.1
    & $\eta(r,\vphi,z)$ & uniform(0)   & 4 &  $R_m=145$  \\
B7    &  0.1      & 10                    & 18         & 9.1
    & $\eta(z)$ & uniform(0)   & 4 &  $R_m=430$ \\
\noalign{\smallskip}\hline
\end{tabular}
\end{flushleft}
\end{table*}

\subsection{Model parameters}

Having combined 3-D SPH code with 3-D induction code, we simulate
the evolution of magnetic-field structure and strength in a model
spiral galaxy. Although there are many factors to affect
velocity-field properties, owing to the limitation of available
computation time, we will focus only upon some characteristic
gas flows.

The model parameters adopted in the computations are summarized
in Table 1. Our models are divided into 2 groups depending on
the initial condition for the magnetic field: axisymmetric and
uniform field configurations. Though more general would be the
superposition of several magnetic modes (Panesar \& Nelson 1992),
these initial conditions are enough to see the possible effects
of non-axisymmetric velocity disturbances, and also the predominant
field configurations revealed in radio observations.
In Table 1, the models $Ai(i=1-4)$
are devoted to the initially axisymmetric magnetic-field configuration.
Models $Bi(i=1-7)$
correspond to  initially uniform (parallel to the Y axis)
magnetic-field structure.
The intensity of the field is $1~\mu G$ in both cases from the
reason that the microgauss order of initial magnetic fields after a disk
formation has been suggested by the observations of damped Ly$\alpha$ clouds
(e.g. Kronberg et al. 1992, Wolfe et al. 1992), and an idea to explain such
observations was given in Lesch and Chiba (1994). Note that since the
present concern
is devoted to linear processes and the resultant structure of magnetic fields,
the actual value of initial field strength does not change our results.
The vertical velocity of gas is forced to be zero in all experiments
for the sake of simplicity.
For the diffusion coefficient $\eta $, we adopt the magnitude resulting
from  the
small-scale  turbulent   gas   motions. Its basic value in a galactic plane is
$\eta _{0}~=~5.7\cdot 10^{27}$~cm$^2$~s$^{-1}$. This rather high value for
$\eta_{0}$ compared to the usually adopted value is to reduce the computational
time for solving the induction equation. We have confirmed from the simulation
with smaller $\eta_{0}$ that the essential results do not change at all.
We note
that this order of magnitude for $\eta_{0}$ is
close to  the  upper  limit  of  the
turbulent diffusion coefficient for  the  interstellar  gas
suggested in the literature (Lesch 1993).

We have introduced the $z$-dependent diffusion coefficient, $\eta(z)$:
the  presence  of  the  active
corona in the galaxy with high temperature $(10^{6}~K)$  and  the  high
turbulent velocity causes the  increase  of $\eta $  in  the  direction
perpendicular to the galactic plane (Camenzind \& Lesch 1994;
Spencer \& Cram 1992).  From  this  reason  we  have  applied
the following form for $\eta$  (Spencer \& Cram 1992):
\begin{equation}
\eta = \left\{
  \begin{array}{ll}
     \eta_0 &  {\rm for}\ z \le H_ {\rm dif} \\
     \eta_0\cdot \exp\left((z-H_ {\rm dif})/h\right) &
                    {\rm for}\ H_{\rm dif} < z \le H_{\rm max} \\
     \eta_{\rm max} & {\rm for}\ z > H_ {\rm max},
  \end{array} \right.
\end{equation}
where $\eta _{0}$ is our basic  value  of  the  diffusion  coefficient,
$H_{\hbox{dif}}$ ($=$~0.4~kpc) is the scale height  of  the  interstellar gas,
and $h$
is the scale length of the increase of $\eta$.  In  our  case, $h~=~500$ pc,
so that $\eta $ increases about 2.5 times that in the  galactic  plane.  In
order  to  stabilize  the  code  near  the  boundaries  in  the vertical
direction, the diffusion coefficient above $H_{\rm max} = 1$ kpc is
constant with a value
$\eta _{\max }~\cong ~1.4\cdot 10^{28}$~cm$^2$~s$^{-1}$.
The average value of $\eta$ adopted above suggests a
magnetic Reynolds number $R_{m}~\cong ~55$ [defined
as $R_{m}=v_{\max } L/ \eta $,
where $v_{\max }$ is the maximum value of the velocity field
(200~km~s$^{-1}$),
and $L$
the radius of the model galaxy (10 kpc)]. Relatively  high
value of $\eta $ causes the characteristic evolution time of magnetic field
less than $10^{9}$~yrs, thereby
allowing us to reduce the actual time of  computation.
However, for one experiment, we  have
adopted the diffusion coefficient ten times smaller than the basic one,
for which
we have applied $\Delta t~=~10^{5}$ yrs and assumed the uniform
magnetic  field
as an initial field configuration. As a result, we
obtain $R_{m}~=~430$ (Table~1,~B7).

In order to take into account the influence of spiral-arm activity
on the
diffusion coefficient, we have adopted the following form for $\eta(r,\vphi)$:
\begin{eqnarray}
\eta(r,\vphi) &=& \fr{\eta_0-\eta_{min}}{2}
 \left[ 1+ \cos \{ 2\vphi + 2\ln (r/r_0) \cot \psi \} \right]
 \nonumber \\
&& + \eta_{min},
\label{eta}
\end{eqnarray}
where $\eta_{min}$, the minimum value of $\eta$, is equal to
$\eta_0 / 10$ in our model. Equation (22) is equivalent
to eq.(14), i.e. we assume the two-armed logarithmic
spirals for $\eta$. Since the turbulent velocity of interstellar
gas, $<v^2>^{1/2}$,
would be high in arm regions due to efficient energy input
from young OB stars and supernovae, the turbulent diffusion,
$\eta \sim <v^2>^{1/2}l$ where $l$ is the correlation length
of turbulence, would be also large (Ko \& Parker 1989). Therefore,
we assume in our simulations that $\eta$ is maximum in arm regions.
Of course, it is not obvious as to whether this assumption is
realistic or not. However, the depolarization of radio
continuum emission preferentially in arm regions may support
this idea (e.g. Horellou et al. 1993; Neininger et al. 1993).
The magnetic Reynolds number is about 178 in this case (B5).
\par
Two experiments (A2 and B6) have been made for the diffusion
coefficient variable
in the galactic plane as well as in in the $z$ direction accordingly to the
above formulaes (21 and 22). The averaged value of $\eta$ gives as a result
$R_m~=~145.$

To avoid numerical instabilities related to the steep
diffusion gradients, $\eta$ has a constant value in the central part
of the disk (to the radius of 2 kpc) in all cases.

\begin{figure*}[hbtp]
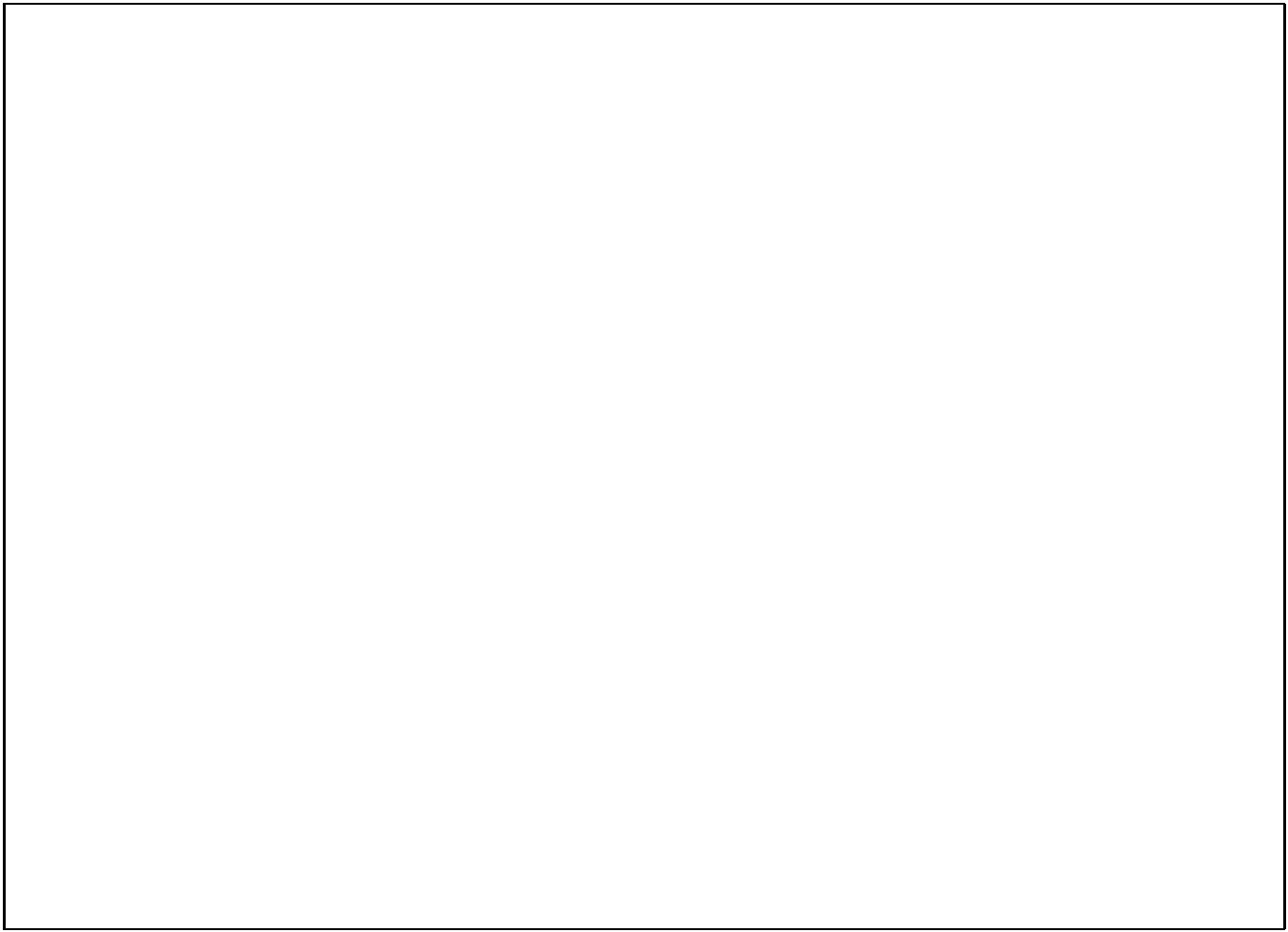

 \picplace{13.0 cm}
\caption[]{\label{fig4} Intensitites of magnetic fields (grayplot)
and magnetic vectors for the model $A1$ at $t=1.1\cdot10^8$ yrs (a)
and at $t=2.2\cdot10^8$ yrs (b), and for the model $A2$ at the
same epochs (c) and (d). The solid lines denote the minima of the
gravitational potential of spiral arms}
\end{figure*}

\section{Results}

\subsection{The $3D$ evolution of magnetic field: the case of initially
an axisymmetric magnetic field}

The calculations with initially a  ring  magnetic  field  structure
have been done for three types of gaseous velocity fields
(Table 1; A):  two-armed spiral disturbances with the pitch angle $\psi $
being $10^{\circ}$ and $20^{\circ}$, and  the  bar  perturbation. By these
experiments, we attempt to clarify  the  net  influence  of
non-axisymmetric disturbance  on  an
axisymmetric  magnetic-field distribution.
In  all  cases,  the
magnetic field follows closely the gas  motions, thus prevailing similar
structures to spirals or a bar at early evolutionary stages. As time
progresses, the
field is expelled out from the inner part of the
disk due to the adopted strong turbulent diffusion.

Fig.~4 presents the grayplots of the intensity and the vectors  of
the magnetic field in the galactic plane,
for the models $A1$ with $\eta(z)$ ($a$ and $b$),
and $A2$
with $\eta(r,\phi,z)$
($c$ and $d$). Fig.~4a
and 4c corresponds to the evolutionary stage of $t~=~1.1\cdot 10^{8}$~yrs,
while
Fig.~4b and 4d for $t~=~2.2\cdot 10^{8}$~yrs.
The  solid  lines denote  the minima  of   the
gravitational  potential of spirals.  For   both   experiments,  the
initially circular field is transformed into the spiral  structure
($a$ and $c$) and then diminishs by the strong diffusion ($b$  and $d$).
The intensity maxima of the magnetic field
are visible slightly  outside the  potential
minima (i.e. down-stream side),
which are close to the regions where the  velocity  shear  is the
highest one. Comparing Fig.~4 with the gas  density  map  for
$\psi~=~10^{\circ}$
(Fig.~3a), it follows that the maxima of the magnetic  field
coincide with those of gas density, though in the  magnetic
field case, the outer magnetic layers of the galactic  disk  are  much  more
dispersed.
\begin{figure}[hbtp]
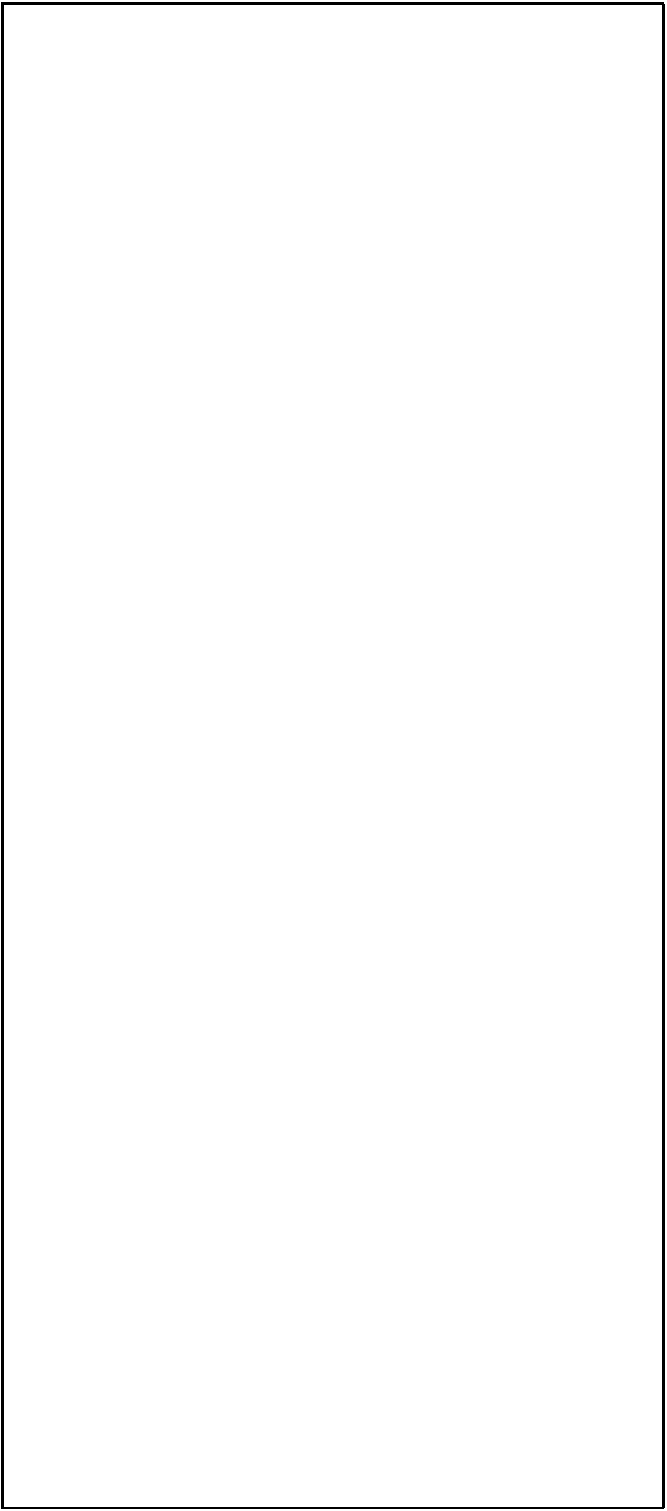

 \picplace{20 cm}
\caption[]{\label{fig5} The grayplots of pitch angle distributions
of magnetic fields for $\psi=10^{\circ}$. The dark (white) areas
denote the positive (negative) pitch angles (the colors range
from -25$^{\circ}$ to 25$^{\circ}$ every 2$^{\circ}$).
The solid lines mark the
gravitational potential minimum.
 (a): $A1$ with $\eta(z)$.
(b): $A2$ with $\eta(r,\varphi,z)$. Fig.5(c) shows the pitch-angles
of velocity fields}
\end{figure}

Fig.~4 also shows the visible  difference  between  the
experiment with the diffusion coefficient $\eta~(z)$ and $\eta~(r,\phi,z)$: the
applied contrast of $\eta $ between the interarm and arm regions  causes
higher magnetic field amplification near  and  inside  the  spiral
arms. It means that  the  variability  of  the  diffusion
coefficient can play an important role in the evolution of galactic magnetic
fields.

One of the most important goals of our simulations  is  to  answer
the  question  how  closely  the  magnetic   field   follows   the
interstellar gas motions. In order to analyze it, we introduce  the
pitch angles of magnetic and velocity fields, defined as the angle between the
field vector and the azimuthal direction
in the galactic  plane.  We note that the obtained distribution of
magnetic pitch angles is not directly compared with observational
results, because the model does  not  take  into
account some observational effects such as the beam smoothing
or the Faraday rotation. Our aim is
to clear the physical connection between magnetic and velocity
fields, which is free from ambiguities in interpretation of observations,
and tremendous number of freedom in parameters depending on galaxy
inclinations,
disk thickness, effects of small-scale magnetic fields, and so on.

Fig.~5 shows the grayplots of the magnetic pitch angles at
evolutionary time $t~=~1.1\cdot 10^{8}$~yrs in the cases of
$\eta ~(z)$ ($A1$) (Fig.~5a) and
$\eta ~(r,\phi ,z)$ ($A2$) (Fig.~5b), as well as, the  pitch-angle
distribution of the velocity field with $\psi ~=~10^{\circ}$~(Fig.~5c).
The dark areas correspond to positive
pitch angles  -  the  vectors  are
directed outward from the disk compared to the azimuthal direction,
while the white ones for negative pitch angles
- the vectors are directed inward.
The  absolute maximum  values of the modelled pitch angles are below
30$^{\circ}$ what is in good agreement with observational data (e.g.
Neininger 1992, Beck 1993).
 The solid line, as in
Fig.~4, shows the minimum of gravitational potential of spirals. It is
easily seen that all three maps for pitch angles are very
similar: the velocity field as well as the magnetic  one  is  directed
outward in front of the potential minimum (i.e. upper-stream side),
then is  aligned  along
the spiral arms, and finally is directed  inward  the  galaxy.  The
highest contrast in the intensities of pitch angle  is  obtained  for
$\eta ~(r,\phi ,z)$, manifesting the greatest evolutionary changes
of magnetic-field structure
due to the gas motions and the diffusion.

\begin{figure*}[hbtp]
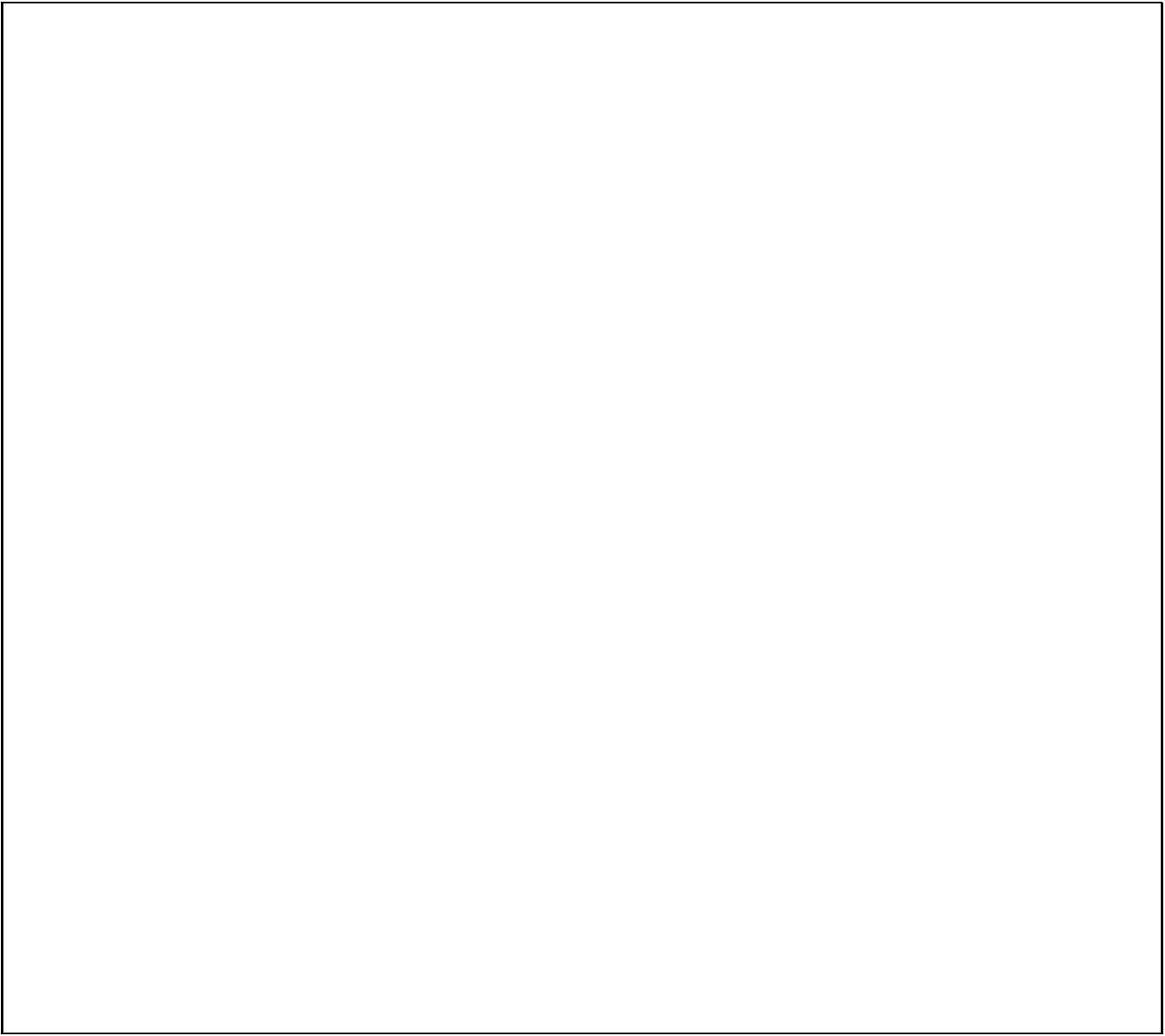

 \picplace{16 cm}
\caption[]{\label{fig6} Results of the model $A3$. (a) and (b)
show the magnetic-field intensity (grayplot) and vectors at
$t=1.1\cdot10^8$ yrs and $t=2.2\cdot10^8$ yrs, respectively. (c) and (d)
show the pitch-angle distributions of magnetic and velocity fields,
respectively. The solid lines denote the gravitational potential minimum}
\end{figure*}

The numerical experiment using spiral velocity fields with
$\psi ~=~20^{\circ}$ (model $A3$) has resulted in the higher
amplification of the
magnetic-field intensity (Fig.~6a and 6b, the legend is the same as in
Fig.~4a and 4b). The maxima are distributed  just  outside  the  potential
minima, more significantly than in the  previous case. As obtained
for $\psi ~=~10^{\circ}$, the magnetic maxima coincide with the gas density
maxima (see Fig.~3b). Fig.~6a is drawn at $t~=~1.1\cdot 10^{8}$~yrs.
When we apply two times longer evolution (Fig.~6b), the magnetic flux
decreases more slowly than in the $A1$ or $A2$ computations.
Fig.~6c and 6d present the pitch angle grayplots for the  magnetic
and velocity fields, respectively. The description is  the  same  as
for Fig.~5. The distribution of both fields is  almost  identical.
The magnetic as well as the velocity field  are  directed  outward
from the disk in front  of  the  spiral  potential,  then are tightly
aligned with the arms, and finally they point inward.
Thus the situation is similar to the $\psi ~=~10^{\circ}$ case,
but now, due to the
higher non-axisymmetricity in the gas flow, the shear is
stronger, causing higher magnetic field amplification.
However the absolute
maximum value is again less than 30$^{\circ}$.

The last case for initially a ring magnetic field is the
experiment with the bar perturbation on  the gaseous velocity  field
(model $A4$; Fig.~7). The magnetic intensity maxima
are not large (Fig.~7a) and they  are  fully  dispersed at the
evolutionary stage of $t~=~2.2\cdot 10^{8}$~yrs (Fig.~7b).
In the central part of the disk,
the strong diffusion sweeps the flux quite significantly. The shear in
the gas velocity is not high so that the amplification of magnetic
field is low. The distribution of the magnetic pitch angles,
which are also similar to those of the
velocity fields, proves that the magnetic field is aligned along the bar and
arms, changing its direction
from outward to inward across the bar and spiral structures.

The magnetic energy density  decreases  in  all  cases  presented,
reaching the lowest value after $5\cdot 10^{8}$~yrs due to the strong
turbulent diffusion.

\subsection{Results for initially a uniform magnetic fields}

The calculations of magnetic field evolution with initially a uniform
(parallel to the $Y$ axis) magnetic field have been performed for four
velocity field configurations (Table 1; models $Bi$): the pure
circular motion with differential rotation, spiral disturbances with
$\psi$ being $10^{\circ}$ and $20^{\circ}$, and the bar perturbation.

For all experiments, the  magnetic
field is wound into the bisymmetric structure, causing the
initial growth of the magnetic energy density (Fig.~8).  As time
progresses, the flux in the inner part of the disk is quickly swept out
by the turbulent diffusion, and the magnetic energy decreases.
Fig.~8 demonstrates the
time evolution of the magnetic energy density $\epsilon $ normalized to  the
initial energy density $\epsilon _{0}$.  The  curves  are  presented  for four
chosen cases with Reynolds number $R_{m}~=~55$ ($Bi(i=1-4)$),
$R_m~=~178$ ($B5$), $R_m=145$
($B6$)
and the case with $R_{m}~=~430 (B7)$.
On the contrary to the experiments with initially a ring magnetic
field, the smallest growth of $\epsilon $ occurs for the spiral  velocity
with $\psi = 10^{\circ}$,
while the largest
one for the bar disturbance (the cases $Bi(i=1-4)$). It follows that for
the flow with
higher non-axisymmetricity, the winding and dispersing  of  the
magnetic field is faster. The case with $\eta (r,\phi,z)$ and $\psi=10^{\circ}$
($B6$)
gives the highest increase of the magnetic energy, mainly
due to the smaller average value of the diffusion coefficient.
The experiment $B5$ with the diffusion coefficient variable in the galactic
plane,
but constant in the $z$ direction (what could have place for galaxies without
the active corona) reveals two times bigger growth of $\epsilon$ and
farther maximum with slowly decreasing slope manifesting sustaining of
the magnetic field longer than the computation time.

\begin{figure}[hbtp]
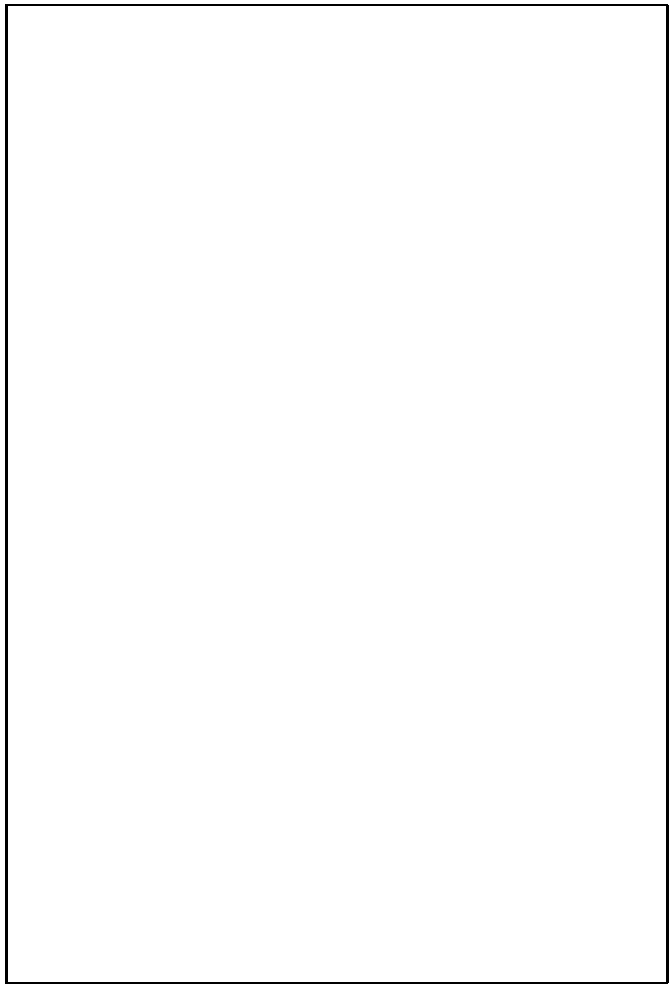

 \picplace{13 cm}
\caption[]{\label{fig7} Distributions of magnetic intensity and
vectors in a bar perturbation (model $A4$) at $t=1.1\cdot10^8$ yrs
(a) and $t=2.2\cdot10^8$ yrs (b)}
\end{figure}

\begin{figure}[hbtp]
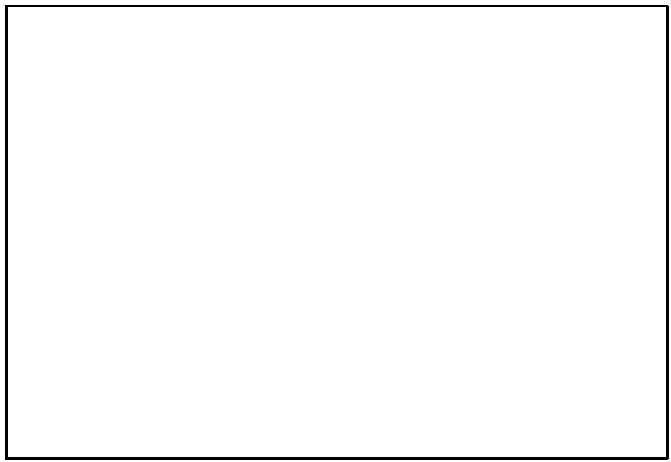

 \picplace{6 cm}
\caption[]{\label{fig8} Time evolution of magnetic energy
density $\epsilon$ normalized to the initial energy density $\epsilon_{0}$
for the models $Bi$. Circ, bar, p20 and p10 denote the velocity
field structures: circular, with bar perturbation, with the pitch
angle of 20$^{\circ}$ and 10$^{\circ}$, respectively}
\end{figure}

The  calculations
for the largest Reynolds number of $R_{m}~=430$ ($B7$) are the most
realistic to
the real galaxy conditions. The growth of the magnetic energy density is
about three times greater than for  the  same  experiment with
$R_{m}~=~55$ ($B2$). The curve in this case does not  decrease  smoothly  but
shows two farther maxima. This fact is well  demonstrated  in  the
maps of the magnetic-field vectors shown in  Fig.~9.  After $10^{8}$~yrs,
the clear bisymmetric structure  is  visible  with  the
slightly dispersed in the center (Fig.~9a).  This  configuration, though more
diffusive, can be still observed at $t~=~2\cdot 10^{8}$~yrs (Fig.~9b) and
$t~=~3.5\cdot 10^{8}$~yrs (Fig.~9c). In  Fig.~9d, the magnetic flux
is completely swept from the central part of the disk,
whereas there is still a kind of ring-like field structure in the regions near
the edges. It means that for experiments  with  smaller  diffusion
coefficient, magnetic field structures will be longer affected by
the non-axisymmetric gas flows.

\begin{figure*}[t]
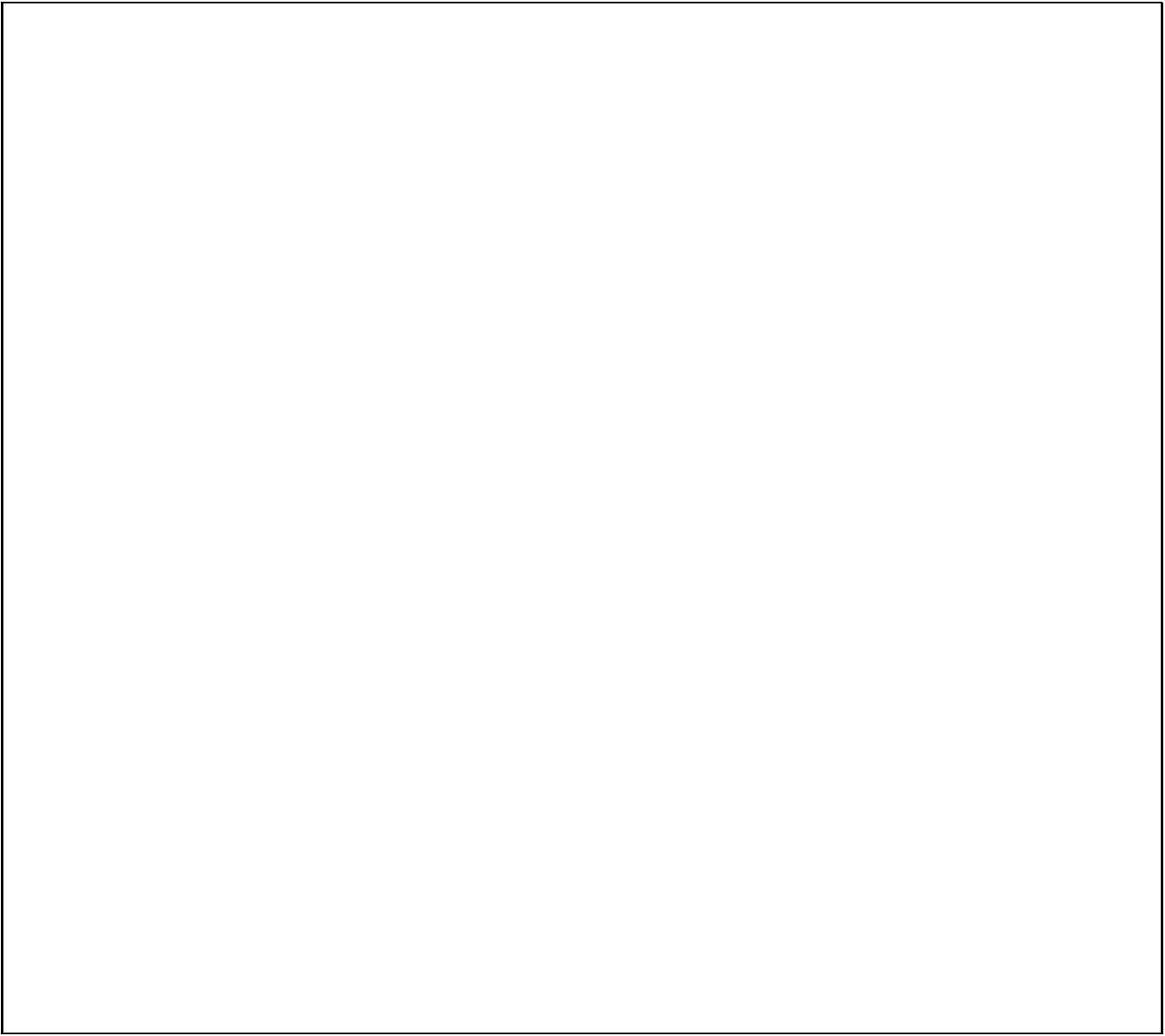

 \picplace{16 cm}
\caption[]{\label{fig9} Distributions of magnetic
vectors for the model $B7$ at 4 evolutionary epochs}
\end{figure*}

\section{Conclusions and discussion}

The evolution of large-scale magnetic field in the galactic disk
affected by non-axisymmetric structures like spirals and bars
has been demonstrated using the 3D numerical simulations.
First, to obtain the realistic velocity fields for calculations
of magnetic fields, the hydrodynamical simulations using SPH
scheme have been performed for three potential disturbances
in an axisymmetric disk: tightly wound spiral with pitch angle
$\psi$ of $10^{\circ}$, open spiral with $\psi=20^{\circ}$
and bar perturbation. Then, the resultant velocity fields of
gas at $\sim 1.6\cdot 10^{8}$ yrs have been used to simulate
the evolution of large-scale galactic magnetic fields.
Our calculations of magnetic fields are devoted to a disk structure
which is well established observationally and theoretically:
some other effects such as vertical gas flows are neglected
because these are tediously increasing parameters for models.
Two basic configurations for the initial magnetic field have been chosen:
axisymmetric (Table 1, $Ai$) and uniform (parallel to the Y axis)
(Table 1, $Bi$). The adopted field configurations are enough to
demonstrate the net effects of non-axisymmetric flow perturbations
on magnetic field structure.

\subsection{Magnetic field structure}

The robust conclusion from our simulations is that
the magnetic field follows closely the interstellar gas motions;

The comparison between magnetic energy and gas density
in the case of initially ring magnetic field ($Ai$)
shows that the maxima of magnetic energy coincide with those of
gas density (Figs. 3, 4, and 6 a and b), and this is also the case for
initially
uniform field. Both maxima are locating slightly outside the minima
of spiral potential (i.e. down-stream side) within the corotation radius
of 9.1 kpc, being the characteristic non-linear response of interstellar
gas (if we consider the effects of magnetic pressure consistently, the
region of density maxima can be broadened, see e.g. Tubbs 1980).

The pitch angle distributions of both velocity and magnetic fields also
indicate the tight correlation between the interstellar gas motion and
magnetic-field structure (Fig.~5a, b and c; Fig~6c and d):
both distributions of pitch angles are very similar even for the longest
evolutionary time. This suggests that the characteristic gas motions
involved in a non-axisymmetric disk are dominant enough
to organize the field structure even in the presence of turbulent dissipation.
The actual spatial variation of field lines of force is summarized as follows:
In front of the spiral arm (i.e. upper-stream side), the orientation of
magnetic field lines is slightly deviating from the arm, with the absolute
value of pitch angle $\psi$ being small or positive $\psi$
(i.e. more leading sense than
the trailing spiral arm itself). {\it In} the arm, the field lines are well
aligned with the arm, i.e. the absolute value of $\psi$ is maximum with the
negative sign. Then, outside the arm, the orientation of field lines is
changing gradually from the trailing sense in the arm to the leading sense
as the gas stream leaves the arm, so that the field crosses with
the next arm again with a finite angle.
Thus, the variation of the magnetic pitch angles with azimuth is
doubly-periodical, and the properties of the pitch-angle variation
are essentially in agreement with the observational
results for NGC6964 (Ehle \& Beck 1993), M51 (Neininger 1992), M83 (Neininger
et al. 1991), and M81 (M.Krause, private communication). We remark that
the definition of pitch angles by radio observers is different from the
usual one, and thus one should be careful when saying the {\it sign} of
pitch angles. Also, our estimate of the field orientation does not include
some observational effects such as the beam smoothing,
Faraday rotation and the effects of
small-scale turbulent fields, so that the observed distribution of the field
lines obtained above may be somewhat different but the general properties
of field structure across the arm may be unchanged.

We have clarified that the actual structure of magnetic field is well
correlated with the optical structure of a galactic disk as radio observations
have revealed. We are thus puzzled how other approaches based on
conventional $\alpha\omega$-dynamos with only an axisymmetric disk can explain
such universal correlation without relying on spiral-arm disturbance.

\subsection{Time evolution of magnetic field energy}

All of our simulated models result in the decay of magnetic energy density
in the absence of field generation mechanisms.
In the cases of initially ring fields ($Ai$), the magnetic energy density,
$\epsilon$, simply decreases with the evolutionary time
due to the weakness of the coupling between convectional motion of gas and
magnetic fields, $|{\bf v}\times{\bf B}| \ll 1$.
As mentioned before, each model
suggests the amplification of magnetic field in the arm due to compressional
and additional shearing motions of gas induced by the arm;
the strongest amplification is obtained for the spiral potential
when the pitch angle $\psi$ of $20^{\circ}$ ($A3$).
On the other hand, the experiments $Bi$ have resulted in
the wound magnetic-field configurations causing the initial
growth of the magnetic energy density $\epsilon$ in the first $10^{8}$ yrs
of evolution (Fig.~8). The fast increase of $\epsilon$ is followed by
quick decaying of its value due to the winding of magnetic lines and
the resultant stimulated effects of diffusion.
The higher non-axisymmetric gas flow causes the lower growth of the magnetic
energy density because of the stronger reconnection of
the magnetic field lines. In two given cases ($B5$ and $B7$),
the longer sustaining of $\epsilon$
is observed: in the case $B5$ with the  diffusion coefficient variable
in the galactic plane and in the experiment $B7$ with the high value of
the magnetic Reynolds number of 430. The case $B5$ has the smaller average
value of $\eta$ and higher $R_m$ (178) than others, so the slower decrease
of $\epsilon$ is understandable.

The anti-dynamo theorem says that any two-dimensional flows result in
the decay of magnetic fields (e.g. Zeldovich et al. 1983). This is due to
the absence of the coupling between the field components along the plane
of flows and the perpendicular field component. No coupling is eventually
followed by the catastrophic reduction of the characteristic scale of
magnetic fields, and thus the dissipation. As is well known,
the so-called dynamo cycle relying on helical motion of turbulence can
introduce the coupling among three components of magnetic fields.
A disadvantage in galactic dynamo theory is that it is not obvious concerning
the actual existence of helical interstellar turbulence from both
theory and observation
({\it cf}. Ferri\`ere 1992; Hanasz \& Lesch 1993; Kaisig et al. 1993). Spencer
\& Cram (1992) have made the important progress
by proposing a new field-generation
mechanism relying on global galactic winds. This approach is however also
subject to the uncertainty as to whether such preferential winds can
be realized in an actual galaxy. It is also worthwhile to note that
their model is based on an axisymmetric disk, and thus suggesting the
generation of non-axisymmetric magnetic fields in order not to break
the anti-dynamo theorem. The emergence of non-axisymmetric fields will
give rise to the existence of non-axisymmetric coupling terms between
velocity and magnetic fields, which will eventually affect the basic
mechanism of field generation from an axisymmetric disk. The general
analysis of such a system is now underway (Spencer, private communication).
On the other hand, observations of edge-on galaxies suggest the mass outflow
from a disk to a halo (e.g. Bloemen 1991), though it is not clear as to whether
such vertical flow is occurring steadily or intermittently.

Therefore, it is necessary to consider the general problem of
magnetic-field evolution containing the realistic vertical gas flow
and the non-axisymmetric velocity disturbance. If the condition for
a dynamo is fulfilled in such a system, the growth of magnetic fields
can be very fast when the efficient amplification due to non-axisymmetric
disturbance, e.g. the wave coupling between density and magnetic waves,
is considered (Chiba \& Lesch 1994), which we have neglected here
by suppressing the propagation effect of spiral arms with azimuth in the
simulations.

\subsection{Effects of azimuthally varied $\eta$}

In the present numerical experiments, we have explored the possibility
that the dissipation effects for magnetic fields are spatially varied
in a galactic plane. Usually the coefficient $\eta$ of turbulent diffusion
has been assumed to be uniform in a plane for galactic dynamos.
Ko \& Parker (1989) have examined the time-dependent dynamo on the assumption
of spatial variability in dynamo efficiency driven by young OB stars and
supernovae in spiral arms. The energy transformation from stellar winds
and explosions to interstellar medium gives rise to high velocity
dispersion of turbulence, thereby enhancing the dissipation of magnetic
fields. Mestel \& Subramanian (1991) considered the enhancement of
helical turbulence ($\alpha$-effects) in spiral arms. Though the exact
expression for $\eta(\varphi)$ is unknown, it is interesting to see
how the effect is non-trivial for magnetic-field evolution.

The model with the diffusion coefficient variable in the galactic
plane ($A2$) has resulted in higher amplification
of magnetic intensity in the spiral arms than in the case for $\eta$
constant in the plane ($A1$). This is well explained by the transport of
magnetic flux from the region where diffusion is high (in our case,
the minima of potential perturbation) to the region where diffusion is weak.
The gas streaming motion across the arm prohibits the leakage of magnetic
flux towards the upper-stream side, so the magnetic intensity is magnified
at the down-stream side of spiral arms. If we consider the generation
of small-scale magnetic fields from such intense turbulence, the total
strength of magnetic fields can be the highest in the arm regions.
Furthermore, the contrast in magnetic pitch angles is also magnified
in this case (Fig.~5b), probably due to the higher coupling between
velocity and magnetic fields in the inter-arm regions where $\eta$ is
assumed to be small.

One more noticeable point for the case of spatially varied $\eta$ is
that the decay of magnetic field energy is made much slower (Fig.~8).
Therefore, more realistic may be the rather slow dissipation of
large-scale magnetic fields compared to the usually voiced value
of dissipation time scale of $\sim 10^8$ yr. Further studies about
this point are necessary before drawing such conclusions.

%

\acknowledgements{We are grateful to Harald Lesch, Marek Urbanik,
Steve Spencer, Rainer Beck, and Elly Berkhuijsen
for their helpful comments and discussions.
KO wishes also to express her gratitude to Zbigniew Kosma and Marian
Soida for their valuable
advice during this work.
Calculations have
been supported by the Forschungszentrum J\"ulich GmbH.
KO thanks the hospitality
of Max-Planck-Institute in Bonn, and MC thanks the Alexander von
Humboldt Foundation for support. This work was partly supported
by the grant from Polish Committee for Scientific Research (KBN),
grant no. PB/0628/P3/93/04.}


\end{document}